\documentclass[]{aastex631}

\usepackage{amsmath}

\begin{document}

\title{Flux-Rope Mediated Turbulent Magnetic Reconnection}

\shorttitle{Flux-Rope Mediated Turbulent Reconnection}
\shortauthors{Russell}

\author[0000-0001-5690-2351]{Alexander J. B. Russell}
\affiliation{School of Mathematics \& Statistics, \\
University of St Andrews, \\ 
St Andrews, KY16 9SS, UK}

\begin{abstract}
We present a new model of magnetic reconnection in the presence of turbulence. The new model differs from the Lazarian-Vishniac turbulent reconnection theory by emphasizing the role of locally coherent magnetic structures, whose existence is shown to be permitted by the properties of magnetic field line separation in turbulent plasma. Local coherence allows storage of magnetic helicity inside the reconnection layer in the form of locally coherent twisted flux ropes. We then introduce the ``Alfvén horizon'' to explain why the global reconnection rate can be governed by locally coherent magnetic field structure instead of by field line wandering, formally extending to 3D the principle that reconnection can be made fast by fragmentation of the global current layer. Coherence is shown to dominate over field line dispersion if the anisotropy of the turbulence at the perpendicular scale matching the thickness of a marginally stable current layer exceeds the aspect ratio of the current layer. Finally, we conjecture that turbulence generated within the reconnection layer may produce a critically balanced state that maintains the system in the flux-rope mediated regime. The new model successfully accounts for the major features of 3D numerical simulations of self-generated turbulent reconnection, including reconnection rates of 0.01 in resistive MHD and 0.1 with collisionless physics.
\end{abstract}


\section{Introduction}\label{sec:intro}

Magnetic reconnection is a fundamental plasma process that enables the rapid reconfiguration of magnetic fields. It is strongly associated with plasma heating, jets and particle acceleration, and it is central to a wide range of explosive astrophysical phenomena including solar flares, coronal heating and auroral substorms. Magnetic reconnection also occurs in the laboratory, for example, during sawtooth crashes and flux rope merging. Introductions to the field can be found in the books by \citet{PriestForbesBook}, \citet{BirnPriestBook} and \citet{YamadaBook}.

The first quantitative model of magnetic reconnection \citep{1958Sweet,1957Parker} examined a laminar steady-state current sheet in resistive MHD, finding that the reconnection rate satisfies 
\begin{equation}\label{eq:rate_SP}
    \frac{V_{rec}^{SP}}{V_A} = S_L^{-1/2},
\end{equation}
{where $V_{rec}$ and $V_A$ are, respectively, the inflow speed and the Alfvén speed upstream of the reconnection layer}.
The right-hand side depends on the Lundquist number,
\begin{equation}\label{eq:Lundquist}
    S_L = \frac{LV_A}{\eta},
\end{equation}
where $L$ is the half length of the reconnection layer and $\eta=1/\mu_0\sigma$ is the magnetic diffusivity. 
It is well known that evaluating Eq~(\ref{eq:rate_SP}) for a typical coronal Lundquist number (say $10^{12}$) yields a reconnection rate ($10^{-6}$) that is vastly smaller than empirical rates of 0.01--0.1 inferred from solar flare observations.

The rate problem with Sweet-Parker reconnection was recognized immediately and various solutions have been explored. The most notable are Petschek reconnection \citep{1964Petschek}, reconnection with collisionless effects  \citep{2001Birn}, turbulent reconnection \citep{1999LV,2009Kowal}, and plasmoid mediated reconnection \citep{2008Lapenta,2009Cassak,2009Bhattacharjee,2010HuangBhattacharjee,2010Uzdensky}. All of the aforementioned models produce acceptably fast reconnection rates. 

Recently, the scientific community has increasingly focused on the cross-scale coupling required to connect the global length scales with the much smaller scales on which field lines reconnect. In this context, the central role of plasmoids in 2D reconnection is well understood \citep{2010HuangBhattacharjee,2010Uzdensky} but it has not yet been conclusively established whether the same mechanism carries over to 3D, or if instead the reconnection rate is determined by fully 3D phenomena such as field line dispersion due to turbulence \citep{1999LV,2009Kowal}. We also allow that different mechanisms may dominate under different conditions. Attempts to resolve these problems have been described by \citet{2023Ji} as the ``third phase of magnetic reconnection research''.

{Plasmoid mediated reconnection and turbulent reconnection are especially relevant to this paper, and we remark at the outset that those theories effectively make opposite assumptions about the structure of the magnetic field. Plasmoid mediated reconnection was originally devised as a 2.5D model and it can therefore be viewed as a theory of how reconnection at high Lundquist numbers works if the magnetic field structure is perfectly coherent. In this limit, coherent magnetic field structures (plasmoids in 2D, or flux ropes in 3D) subdivide the global current layer into short segments that reconnect at a fast rate. In contrast, the turbulent reconnection theory of \citet{1999LV} explains how reconnection at high Lundquist numbers works if field line dispersion dominates, determining the width of the reconnection outflow and hence the reconnection rate. The quantitative theories of these mechanisms are recapped in Appendix \ref{app}. The main goal of this paper is to explore how reconnection works in the middle ground where the magnetic field inside the reconnection layer exhibits coherence over local trace distances and field line dispersion over longer trace distances.}

In the study of solar flares, there is a long-standing association between turbulence and magnetic reconnection. Specifically, the well documented nonthermal broadening of hot spectral lines is widely considered to be direct evidence of turbulent motions. A review covering historical and modern observations and their interpretation has been given by \citet{2024Russell}{, which} also explores the generation mechanisms of turbulence in solar flares and its connection to reconnection. 
Focusing here on recent observational results, space instruments have inferred unresolved motions at hundreds of km s$^{-1}$ in the above-the-loop region of solar flares \citep{2014Doschek, 2017Kontar, 2021Stores, 2023Shen} and in post-CME plasma sheets \citep{2002Ciaravella,2008CiaravellaRaymond,2008Bemporad,2013Susino,2018Warren,2018Li,2018Cheng,2020French}.
The closeness of the relationship between turbulence and reconnection in solar flares is further demonstrated by the detection of turbulence tens of seconds before impulsive phenomena: the papers by \citet{2018Jeffrey} and \citet{2020ChittaLazarian} provide modern examples, also see earlier evidence on timings by \citet{1984Antonucci}, \citet{1998Alexander}, \cite{2001Harra} and \citet{2001Ranns}. Insight into why turbulence and fast reconnection develop side-by-side has recently come from work by \citet{2021French} and \cite{2021WyperPontin}, who studied the substructure of flare ribbons around the onset of the impulsive phase. Both studies concluded that fast reconnection begins after a coronal current sheet becomes unstable to tearing instability which develops nonlinearly into turbulence, with \citet{2021French} showing that ribbon substructure grows fastest at a key scale then spreads to larger and smaller scales.

The scenario of tearing instability developing into turbulence as reconnection becomes fast is also supported by direct numerical simulations. While the first numerical investigations of turbulent reconnection were driven by a forcing term added to the momentum equation \citep{2009Kowal}, the frontier of high performance computing has recently crossed a threshold beyond which turbulence develops by itself within 3D magnetic reconnection simulations. Turbulence is self-generated when the scale separation exceeds a critical threshold (canonically $S_c = 10^4$ in resistive MHD) above which an initial Sweet-Parker current layer is unstable to a fast tearing instability \citep{1963Furth,1978Bulanov,1986Biskamp,2007Loureiro,2012Baalrud}. When $S_L\gg S_c$, the tearing instability can operate recursively, as proposed by \citet{2001ShibataTanuma}. In 3D, the ensuing nonlinear evolution generates and sustains turbulence within the reconnection layer. Self-generated turbulent reconnection (SGTR) simulations of this type have now been reported by many investigators, using both collisionless codes \citep{2011Daughton,2014Daughton,2013Liu,2013Pritchett,2013Nakamura,2017Nakamura,2015Dahlin,2017Dahlin,2018Le,2019Stanier,2019Li,2021AgudeloRueda,2021Zhang} and resistive MHD codes \citep{2015Oishi,2016HB,2016Striani,2017Beresnyak,2017Kowal,2020Kowal,2020Yang,2022Beg}.

A current dichotomy is that SGTR simulations exhibit many features anticipated by the \citet{1999LV} theory of turbulent reconnection, including fast reconnection rates, power law spectra and dispersion of magnetic field lines. However, they also exhibit features that are strongly reminiscent of 2D plasmoid mediated reconnection, such as the presence of flux ropes in the reconnection layer and an excellent match of the reconnection rate \citep{2014Daughton,2016HB,2022Beg}. The present paper aims to develop the theory of 3D magnetic reconnection with self-generated turbulence by addressing the roles of locally coherent magnetic strutures and magnetic helicity. We propose a new conceptual model that reconciles aspects of the 3D Lazarian-Vishniac turbulent reconnection theory and 2D plasmoid mediated reconnection. The new model successfully accounts for the main features seen in SGTR simulations that reach a statistically stationary state, including the reconnection rate.

The article is structured as follows. Section \ref{sec:disp} investigates the dispersion of magnetic field lines in turbulent plasma, showing that while field line wandering dominates over long trace distances, magnetic field structures are coherent for shorter trace distances. The nature of the coherent structures is investigated in Section~\ref{sec:helicity}, where we propose that conservation of magnetic helicity provides a fundamental basis for the presence of locally coherent flux ropes inside the reconnection layer, even at extremely high Lundquist numbers, and consider their 3D interactions. In Section~\ref{sec:horizon}, we introduce the concept of the Alfvén horizon and using this to show that coherent structures can govern the global reconnection rate as opposed to the field line dispersion. The paper concludes with a discussion in Section~\ref{sec:disc} and summary in Section~\ref{sec:summary}. The articles by \citet{1999LV} and \citet{2022Beg} are referred to frequently, so we abbreviate them as LV99 and BRH22.

\section{Magnetic Field Line Separation in Turbulent Plasma}\label{sec:disp}

We begin by investigating the dispersion of magnetic field lines in  turbulent plasma, exploring how the mean square separation between field line pairs $\left<\delta^2\right>$ depends on the trace distance $s$ and the initial separation $\delta_0$. The value of $\left<\delta^2\right>$ tracks the cross-sectional area over which field lines have been dispersed, while the root mean square (rms) separation $\left<\delta^2\right>^{1/2}$ measures the average distance between field line pairs. Our modeling approach follows LV99 with the important difference that we treat the initial separation distance between field line pairs without setting it to zero.

\subsection{Derivation}\label{sec:disp:derivation}

We consider the average behavior over many field line pairs, making a continuum approach suitable, and adopt Eq~(4) of LV99,
\begin{equation}\label{eq:dmss}
\frac{d \left<\delta^2\right>}{d s} = k_\parallel \left<\delta^2\right>,
\end{equation}
where $k_\parallel^{-1}$ is the characteristic length scale governing scattering of magnetic field lines.  If $k_\parallel$ is independent of $\left<\delta^2\right>$ then Eq~(\ref{eq:dmss}) yields the exponentiation relation $\left<\delta^2\right> = \delta_0^2 \exp(k_\parallel s)$.

When field lines are scattered by turbulence, one must take into account that $k_\parallel$ for a magnetic eddy depends on $k_\perp$. In this paper, we consider scaling relations of the form
\begin{equation}\label{eq:kpar_scaling}
    k_\parallel = C k_\perp^a,
\end{equation}
where the proportionality factor $C$ and the index $a$ are properties of the turbulence.
Equation~(\ref{eq:kpar_scaling}) is equivalent to a scale-dependent anisotropy relation 
\begin{equation}
    \frac{k_\parallel}{k_\perp} = C k_\perp^{a-1}.
\end{equation}

For MHD turbulence, one typically has $0<a\leq 1$. If $0<a<1$ then the magnetic eddies become increasingly elongated along the background magnetic field with decreasing perpendicular scale, consistent with the filamentation that is characteristic of magnetized plasmas, i.e. the dynamics become increasingly 2.5D at smaller scales. The \citet{1995GS} turbulence adopted by LV99 satisfies Eq~(\ref{eq:kpar_scaling}) with $a=2/3$ and $C = l_{inj}^{a-1}(v_{inj}/V_A)^{4/3}$, where $v_{inj}$ is the velocity amplitude at injection scale $l_{inj}$. Direct numerical simulations of self-generated reconnection turbulence by \citet{2017Kowal} support the $a=2/3$ scaling. In the special case of $a=1$, the anisotropy is scale-independent, as was reported for the SGTR simulation by \citet{2016HB}.

Using Eq~(\ref{eq:kpar_scaling}) to substitute for $k_\parallel$ in Eq~(\ref{eq:dmss}), and making the identification $k_\perp=\left<\delta^2\right>^{-1/2}$, we obtain
\begin{equation}\label{eq:disp:ode}
    \frac{d \left<\delta^2\right>}{d s} = C \left<\delta^2\right>^{1-a/2},
\end{equation}
which is a separable ordinary differential equation for $\left<\delta^2\right>$. Provided that $a\neq0$ (avoiding the exponentiation case dealt with separately above) the solution is
\begin{equation}\label{eq:super}
    \left<\delta^2\right> = \left( \delta_0^a+ \gamma s \right)^{2/a},
\end{equation}
where $\gamma = Ca/2$. When analyzing a numerical simulation or an experiment, it may often be easier to obtain $\gamma$ by fitting Eq~(\ref{eq:super}) than to infer it indirectly from $a$ and $C$.
The $\delta_0$ that appears in Eq~(\ref{eq:super}) comes from the constant of integration, which was not included by LV99. Its inclusion is central to the present paper.

\subsection{Richardson Dispersion}\label{sec:disp:richardson}

For infinitesimal $\delta_0\to0$ or large $s\to\infty$, Eq~(\ref{eq:super}) behaves asymptotically as
\begin{equation}\label{eq:Richardson}
    \left<\delta^2\right> \approx \left(\gamma s\right)^{2/a}.
\end{equation}
In this limit, the average separation of field line pairs is fully determined by the trace distance $s$ and the turbulence properties $\gamma$ and $a$, whereas the original separation $\delta_0$ has been forgotten. The $\left<\delta^2\right> \sim s^{2/a}$ scaling in Eq~(\ref{eq:Richardson}) implies that turbulence with $a<2$ disperses magnetic field lines faster than Brownian diffusion, which has $\left<\delta^2\right>\sim s$. If $a<1$ such that small-scale magnetic eddies have greater $k_\perp/k_\parallel$ than large-scale eddies, then the rms separation $\left<\delta^2\right>^{1/2}$ increases superlinearly with respect to $s$.

In the specific case of the \citet{1995GS} scalings, Eq~(\ref{eq:Richardson}) yields
\begin{equation}\label{eq:RLV}
    \left<\delta^2\right> \approx \left(\frac{v_{inj}}{V_A}\right)^{4} \frac{\left(3s\right)^{3}}{l_{inj}},
\end{equation}
which recovers the Richardson dispersion scaling $\left<\delta^2\right> \sim s^3$ obtained by LV99.

LV99 noted the assumptions used to derive Eq~(\ref{eq:disp:ode}) eventually fail for $s\to\infty$, since the strong turbulence scalings are eventually replaced by diffusion at a maximal rate (the $L>l$ case in Appendix \ref{app:LV}). Under these conditions, the right hand side of Eq~(\ref{eq:dmss}) is replaced with a constant, thus Richardson-type dispersion eventually gives way to Brownian diffusion with $\left<\delta^2\right> \sim s$. In a bounded domain, $<\delta^2>^{1/2}$ is also bounded by the domain size.

\subsection{Local Coherence}\label{sec:disp:local}

The most important difference between Eq~(\ref{eq:super}) and the corresponding derivations in LV99 and Appendix \ref{app:LV} is the inclusion of the initial separation distance. That is to say, we have retained the constant of integration introduced by integrating Eq~(\ref{eq:disp:ode}), which was dropped by LV99. Examining the full solution given by Eq~(\ref{eq:super}), the Richardson-type dispersion regime is only valid for $\gamma s \gg \delta_0^a$. Earlier in the field line tracing, $\gamma s \ll \delta_0^a$ gives $\left<\delta^2\right>^{1/2}\approx \delta_0$, which implies that the field line separation is determined by its initial value. This result allows the presence of locally coherent magnetic field structures, which we propose can have an important role in turbulent reconnection.

\begin{figure}
\centering
\includegraphics[width=10cm]{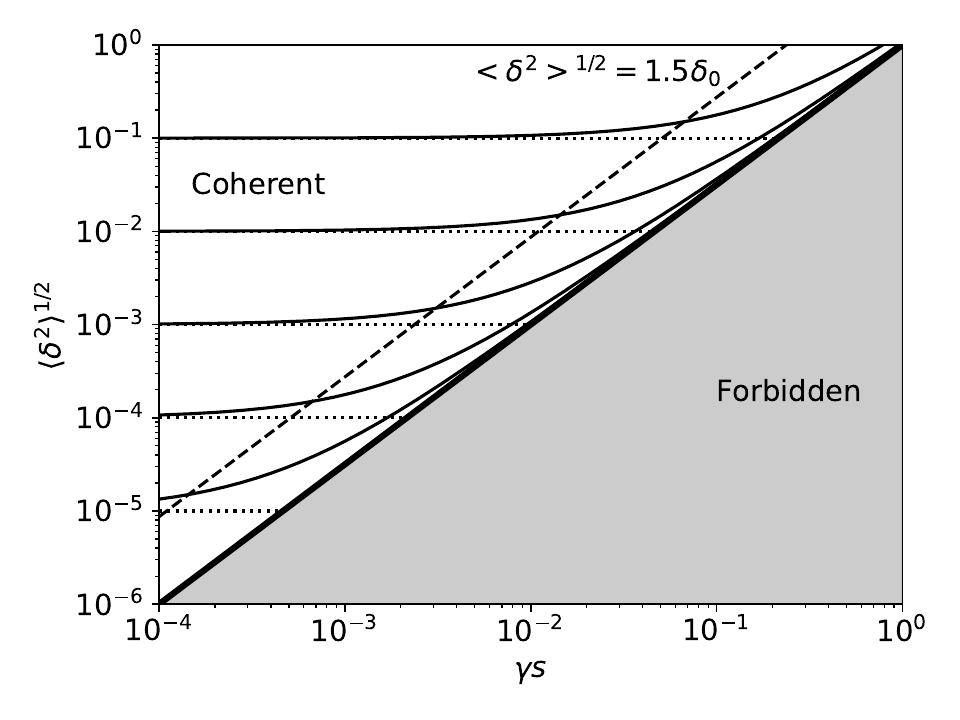}
\caption{Root mean square (rms) field line separation $\langle \delta^2 \rangle^{1/2} = (\delta_0^a+\gamma s)^{1/a}$, showing dependence on trace distance $s$ and initial separation $\delta_0$, plotted for the \citet{1995GS} scaling index $a=2/3$. The thick black line shows the Richardson dispersion relation $\langle \delta^2 \rangle^{1/2} = (\gamma s)^{1/a}$ obtained for $\delta_0=0$. The gray region is forbidden because rms field line separations cannot be smaller than the solution for $\delta_0=0$. Solution curves are shown for $\delta_0 \in \left\{ 10^{-5}, 10^{-4}, 10^{-3}, 10^{-2}, 10^{-1} \right\}$ (thinner solid lines). Initially, $\langle \delta^2 \rangle^{1/2} \approx \delta_0$, which allows the existence of locally coherent magnetic field structure in turbulent plasma. The scale-dependent transition between coherence and dispersion is highlighted by the dashed line, which curves cross when $\langle \delta^2 \rangle^{1/2} = 1.5 \delta_0$.}
\label{fig:dispersion}
\end{figure}

Figure~\ref{fig:dispersion} plots $\langle \delta^2 \rangle^{1/2}$ based on Eq~(\ref{eq:super}), setting $a=2/3$ consistent with \citet{1995GS}, LV99 and \citet{2017Kowal}. The thick black line shows the Richardson dispersion obtained by setting $\delta_0=0$, which is central to the LV99 theory of turbulent reconnection. The filled gray region is forbidden because Eq~(\ref{eq:super}) implies $\langle \delta^2 \rangle^{1/2} \geq (\gamma s)^{1/a}$, with equality only for $\delta_0=0$. Figure~\ref{fig:dispersion} also plots solutions for various non-zero initial separations: these curves are initially horizontal, corresponding to the result that turbulence initially has a negligible effect on the separation. Dispersion only takes over once the dotted horizontal line given by $\langle \delta^2 \rangle^{1/2} \approx \delta_0$ nears the thick solid Richardson dispersion line given by Eq~(\ref{eq:Richardson}).

Inspecting Eq~(\ref{eq:super}), the transition between coherence and dispersion happens as $s$ approaches
\begin{equation}\label{eq:s_star}
    s^* = \frac{\delta_0^a}{\gamma}.
\end{equation}
It is evident that the trace distance over which magnetic structure remains coherent is scale-dependent, since the value of $s^*$ scales as $\delta_0^a$. Mathematically, Eq~(\ref{eq:s_star}) matches the two terms inside the bracket on the right hand side of Eq~(\ref{eq:super}). Physically, dispersion only affects a given length scale once a field line pair with infinitesimal initial separation has been dispersed to the scale of interest. This implies that structure at any chosen scale mixes out after structure at all smaller scales, provided that $a>0$.

Equation~(\ref{eq:s_star}) can be related to the parallel wavenumber of the turbulence at the perpendicular scale $\delta_0$. Using Eq~(\ref{eq:kpar_scaling}), if $k_\perp$ is set to $\delta_0^{-1}$  then $k_{\parallel}(\delta_0) = C \delta_0^{-a}$ and Eq~(\ref{eq:s_star}) implies
\begin{equation}\label{eq:s_star_kpar}
    s^* = \frac{2}{a k_{\parallel}(\delta_0)}.
\end{equation}
Thus, a pair of field lines must be traced a distance comparable to a few times the parallel length scale of the turbulence at the perpendicular scale of their initial separation before field line dispersion by the turbulence becomes significant. 

{This mathemetical result corresponds to a set of physical principles that: (i) field lines are scattered by resonant magnetic eddies whose transverse scale matches the field line separation, and (ii) coherence transitions to dispersion when the effects of several magnetic eddies have been compounded. Furthermore, if the parallel scale of magnetic eddies in MHD turbulence increases with their perpendicular scale (corresponding to $a>0$), the principle that field line pairs transition from coherence to dispersion when traced through a certain number of resonant eddies implies that structures with greater transverse scales mix out after stuctures with smaller transverse scales.}

When $s=s^*$, the rms field line separation satisfies $\langle \delta^2 \rangle^{1/2} = 2^{1/a} \delta_0$. More generally, one might consider dispersion to be significant once $\langle \delta^2 \rangle^{1/2}/\delta_0$ exceeds a value $f$ other than $2^{1/a}$. Using Eq~(\ref{eq:super}), this general threshold is crossed at $s=s^\dagger$, where
\begin{equation}
    s^\dagger = (f^a-1)s^*.
\end{equation}
We consider a value of $f=1.5$ to be a useful indication that the field lines are beginning to separate. Curves in Figure~\ref{fig:dispersion} reach this threshold when they intersect the dashed line.

\subsection{Section Summary}
Magnetic fields in turbulent plasmas exhibit both dispersion and local coherence. Richardson-type dispersion occurs for longer trace distances and we note that this phenomenon is at the heart of the LV99 model of turbulent reconnection. However, by retaining a constant of integration, we have shown that the field line separation model solved by LV99 also exhibits local coherence, meaning that field line separations are unaffected by dispersion until they have been traced a few times the parallel scale of the turbulence evaluated at the perpendicular scale matching the initial field line separation, i.e. through several resonant magnetic eddies. We propose that the property of local coherence should also be taken into account in the theory of turbulent magnetic reconnection.

\section{Magnetic Helicity and Flux Ropes}\label{sec:helicity}

Having shown that turbulence permits the existence of locally coherent magnetic field structures, we now examine what form these structures should take inside a turbulent reconnection layer. 
{In this section, we propose that the reconnection layer typically contains locally coherent flux ropes and argue that their presence is a robust consequence of conservation of magnetic helicity.} 

\begin{figure}
\centering
\includegraphics[width=13cm]{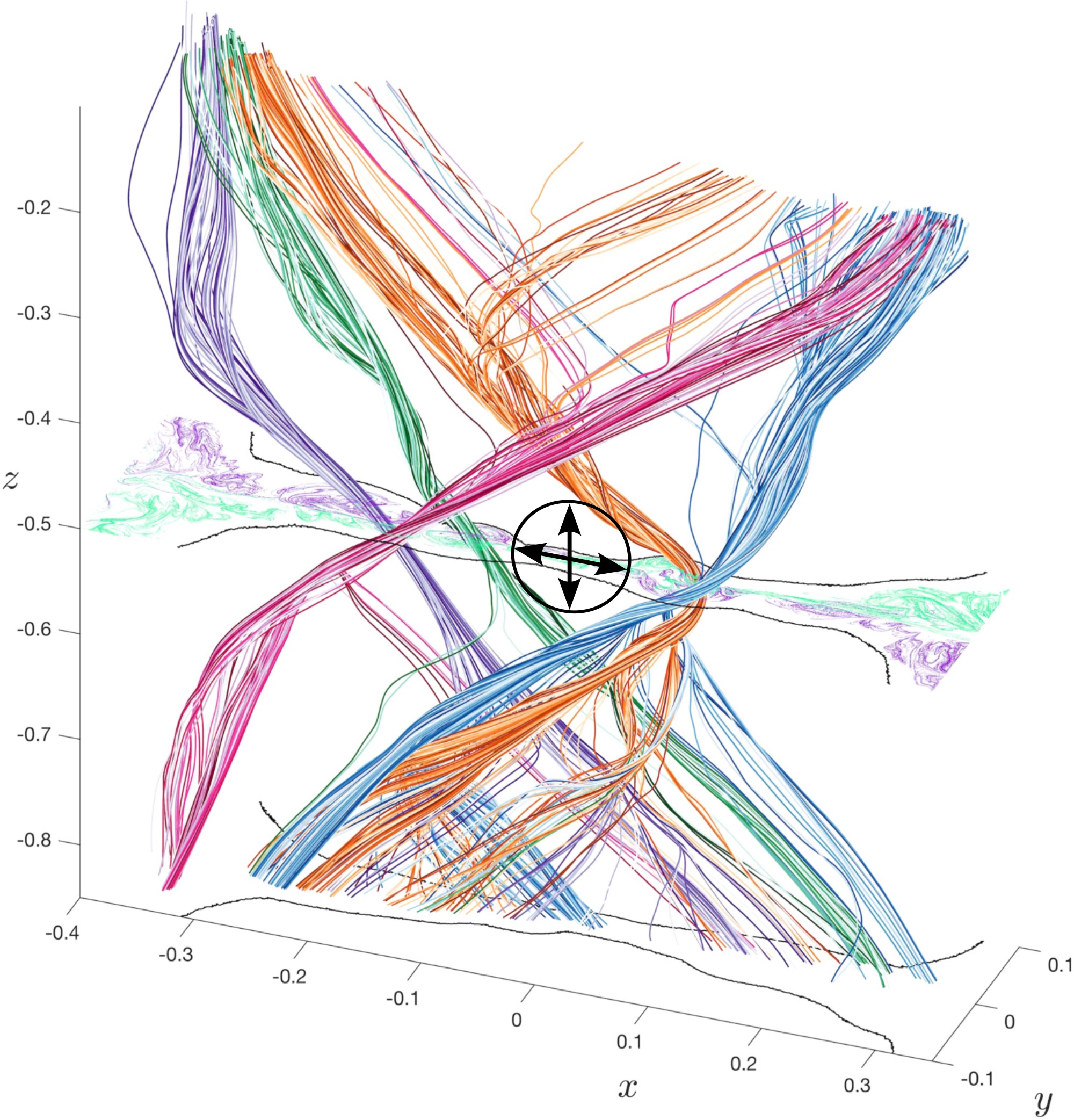}
\caption{Magnetic field lines in an MHD simulation of self-generated turbulent reconnection by \citet{2022Beg}. The black curves on $z=-0.5$ and $z=-1.0$ show where these planes intersect the stochastic spearatrices, which are the upstream boundaries of the turbulent reconnection layer. Magnetic field lines (multicolored) are traced inside the reconnection layer from seed points placed on the $z = -0.5$ plane. Over short trace distances, the field lines form locally coherent oblique twisted flux ropes. Over longer trace distances, dispersion dominates such that field lines from different flux ropes are highly intermixed by the time they reach the $z$ boundaries. Topological structure on the $z=-0.5$ cut is shown using maximizing ridges of the forward (green) and backward (purple) finite time Lyapunov exponent. The horizontal black arrows indicate a representative length of a microscopic current sheet between flux ropes. The black circle indicates the corresponding Alfvén horizon and the vertical black arrows show the maximum $z$ distance inside the Alfvén horizon. The magnetic flux ropes are highly coherent when traced over the horizon distance. Adapted from \citet{2022Beg} under a CC-BY 4.0 license.} \label{fig:beg}
\end{figure}

This section is motivated by Figure~\ref{fig:beg}, which visualizes the magnetic field inside the SGTR layer simulated by BRH22. Bundles of magnetic field lines inside the reconnection layer are traced from the midplane, using different colors for different bundles. Over shorter trace distances, each bundle of field lines forms a coherent oblique twisted flux rope. Over larger trace distances, field line dispersion takes over, such that the magnetic field becomes highly intermixed before it reaches the $z$ boundaries, as expected from Section~\ref{sec:disp}.

More generally, the proposition that turbulence generated in a reconnection layer contains locally coherent flux ropes is reasonable given that: (i) flux ropes have been extensively linked to reconnection at the Earth's dayside magnetopause \citep{1979RussellElphic,1985LeeFu}, in Earth's magnetotail \citep{2004Zong,2008Chen} and in solar flares \citep{2012Takasao,2016Takasao,2018Cheng}; (ii) flux ropes feature in 2D and 3D tearing instabilities \citep{1963Furth,2007Loureiro,2012Baalrud}; and (iii) their presence has been widely noted in SGTR simulations including \citet{2011Daughton}, \citet{2016HB} and BRH22.

At the same time, one should be wary of extrapolating from current simulations to greater Lundquist numbers that are currently inaccessible. Reasonable concerns include the possibility of an undiscovered ``phase transtition'' in reconnection behaviour, similar to the well known one at $S_L\sim 10^4$, or that flux ropes might not survive in turbulence that is more fully developed than in current simulations. We therefore consider it important to place the presence of locally coherent flux ropes on a foundation of deduction from first principles.

In this spirit, the rest of this section argues that flux ropes are a robust feature of magnetic reconnection with self-generated turbulence because: (a) local coherence is built into magnetic field line separations as shown in Section~\ref{sec:disp}; and (b) the production of flux ropes is driven by helicity conservation. Therefore the flux ropes cannot be eliminated by increasing the Lundquist number or making the system more turbulent. The only way to remove them while conserving magnetic helicity is to reduce the absolute value of the helicity towards zero, as in the important case where the magnetic shear angle approaches 180 degrees.

\subsection{Magnetic Helicity and Formation of Flux Ropes}

Magnetic helicity is an important topological invariant that quantifies the total linking of magnetic flux \citep{1969Moffatt}. In ideal MHD, the magnetic helicity is exactly conserved as a consequence of Alfvén's theorem. It is also highly conserved during magnetic reconnection at high Lundquist numbers, which is the basis of Taylor relaxation \citep{1974Taylor,1986Taylor,1984Berger,2015Russell,2016Pontin,2021Yeates}.

In magnetically closed domains, the magnetic helicity is gauge-invariant and can be calculated using the volume integral
\begin{equation}\label{eq:H}
    H = \int_V \boldsymbol{A}\cdot\boldsymbol{B}\, d^3x,
\end{equation}
where $\boldsymbol{A}$ a magnetic vector potential such that $\boldsymbol{B} = \nabla\times\boldsymbol{A}$ \citep{1958Woltjer}.
Helicity can also be extended to magnetically open domains, either using the relative magnetic helicity, which is gauge-invariant but depends on a choice of reference magnetic field \citep{1984BergerField,1985FinnAntonsen}, or by restricting the gauge of the vector potential to ensure that the helicity integral has a geometrical meaning \citep{2014PriorYeates,2018BergerHornig,2020PriorMacTaggart,2023Xiao}. The relationship between magnetic helicity and current helicity has recently been discussed by \citet{2019Russell}.

SGTR simulations are often initialized with a 2.5D magnetic field of the form $\boldsymbol{B}(x,y)$, where $z$ is the invariant direction. When $B_{z}\neq0$, there is typically a net linking of magnetic flux so the magnetic helicity is non-zero, which is the case we consider in this manuscript.

In the special case of a purely 2D initial magnetic field with $B_{z}=0$, a vector potential of the form $\boldsymbol{A}=A(x,y)\boldsymbol{e}_z$ can be found giving $H=0$ from Eq~(\ref{eq:H}). This zero helicity case corresponds to reconnection with a magnetic shear angle of 180 degrees. It is excluded from our present reconnection model but it will be discussed again in Section~\ref{sec:disc:limits}.

\begin{figure}
\centering
\includegraphics[width=13cm]{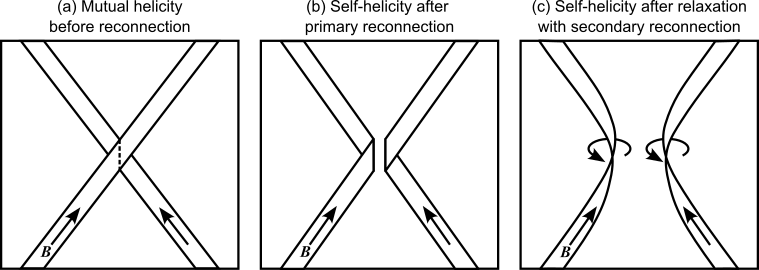}
\caption{Conversion of mutual helicity to self helicity during reconnection of a pair of flux ribbons, based on an original cartoon by \citet{1989WrightBerger}. The left panel (a) shows the original pair of flux ribbons. These cross which gives them mutual helicity. In this cartoon, magnetic reconnection is treated heuristically by cutting the ribbons along the dashed line in (a) and rejoining them as shown in the middle panel (b). The newly created flux ribbons no longer cross so the mutual helicity has been lost and this is compensated by the new flux ribbons having self helicity. The right panel (c) shows the new ribbons after the self-helicity has relaxed into twist distributed along part of the ribbons. For this configuration, both new ribbons have a right handed twist.} \label{fig:ribbons}
\end{figure}

The total magnetic helicity can be decomposed as the sum of self and mutual helicities \citep{1984BergerField}. Typically, when a pair of flux tubes reconnects their mutual helicity changes and the self helicity must therefore change by an equal and opposite amount. At high Lundquist numbers, this process produces twisted flux ropes, as can be illuminated using the following heuristic arguments adapted from \citet{1989WrightBerger} \citep[also see][]{1985LeeFu,2019Wright}.

Figure~\ref{fig:ribbons}(a) shows a pair of flux ribbons on opposite sides of the reconnection layer. They cross each other and therefore have non-zero mutual helicity. In this scientific cartoon, reconnection cuts and rejoins the ribbons to make the new ones in panel (b). The new ribbons do not cross, so there has been a loss of mutual helicity, which has been converted to the self-helicity of the ribbons. At low Lundquist numbers, i.e. in Sweet-Parker reconnection, the magnetic helicity inside the reconnection layer remains in the form shown in Figure~\ref{fig:ribbons}(b). At higher Lundquist numbers, secondary reconnections enable relaxation of the self-helicity into twist distributed along part of the flux ribbon. (In tearing instabilities the multiple reconnections occur simulataneously). Finally, as discussed by \citet{1985LeeFu} and \citet{1989WrightBerger}, multiple reconnections further increase the turns of twist in the new flux ribbons compared to the case sketched here.

The conclusion from these first-principles arguments is that locally coherent flux ropes are a robust consequence of conservation of magnetic helicity. When turbulence is taken into account, the flux ropes lose their identity over longer distances as their field lines become subject to Richardson-type dispersion. Applying the coherence length to a flux rope, the $s^\star$ defined in Eq~(\ref{eq:s_star}) should be evaluated with $\delta_0$ set to the flux rope's diameter. Thus, larger diameter flux ropes remain coherent for longer trace distances, compared to smaller diameter flux ropes.However, coherence only needs to hold locally to store magnetic helicity. These magnetic field properties, inferred from first principles, agree well with those seen in Figure~\ref{fig:beg}.

\subsection{Flux Rope Interactions}

Having explored the reasons why locally coherent twisted flux ropes exist inside the reconnection layer, we now consider how they interact. Previous work on 3D flux rope interactions by \citet{2001Linton} found that flux ropes of the same handedness interact in three ways depending on the contact angle (also see \citet{1996LauFinn} and \citet{1999Kondrashov}). These pairwise interactions operate alongside the dynamics of individual flux ropes, such as kink instabilities which may also play a role in the generation and sustaining of the turbulence \citep{1992Dahlburg,2022Beg}. While the flux ropes we consider fray over longer trace distances, the basic flux rope interactions should provide a useful guide to how shorter sections of locally coherent flux ropes interact. 

The 3D counterpart of 2D plasmoid coalescence is the merge interaction. Merging occurs when the magnetic fields at the edge of the flux ropes are roughly anti-parallel, and Figure~4 of \citet{2001Linton} shows that flux ropes of the same handedness merge for a large range of contact angles. The resulting coalescence starts at the contact point and propagates along the pair of flux ropes as the magnetic tension of the reconnected magnetic field draws a greater length of them together. Applying what is known about the statistical theory of 2D plasmoids \citep{2010Fermo,2010Uzdensky,2012HuangBhattacharjee}, it is likely that 3D merging plays an important role in the growth of large flux ropes such as the ones evident in Figure~\ref{fig:beg}.

When the magnetic fields in the envelopes of the flux ropes are close to parallel, the flux ropes bounce instead of reconnecting. The bounce phenomenon does not change the properties of the flux ropes and it may allow oblique flux ropes in the different sides of the reconnection layer to undergo relatively little reconnection. 

It is also possible for flux ropes to tunnel, which has no 2D counterpart. In this case, field lines that are nearly orthogonal reconnect twice, causing the flux ropes to pass through each other \citep{1997Dahlburg,2001Linton}. The change of positions implies a change in mutual helicity that should be compensated by the self helicity of the flux ropes, hence the interaction changes the twists. According to \citet{1997Dahlburg}, the tunneling interaction requires that the flux ropes have pitch angles significantly greater than those seen in Figure~\ref{fig:beg}, so it is unlikely to be common within turbulent reconnection although it may occur in special cases. 

\subsection{Section Summary}

Turbulent magnetic reconnection must obey laws governing the magnetic helicity, which were not taken into account in previous models such as LV99. At high Lundquist numbers, helicity conservation combined with local coherence of magnetic field lines inside the turbulent reconnection layer creates twisted flux ropes that are the 3D counterpart of 2D plasmoids. These will be a robust feature when the helicity of the reconnecting magnetic field is non-zero, as is the case for many simulations of SGTR. Unlike 2D plasmoids, the 3D flux ropes are typically oblique and they fray as their constituent field lines are dispersed for longer trace distances. Based on previous studies, we expect the most common pairwise flux rope interactions to be merging and bouncing. Merging can help locally coherent flux ropes grow to a considerable size, while the bounce interaction can suppress reconnection between oblique flux ropes in the different sides of the reconnection layer.

\section{Coherence Versus Dispersion}\label{sec:horizon}

The previous sections have demonstrated that turbulent reconnection layers contain locally coherent flux ropes that store magnetic helicity (Section~\ref{sec:helicity}), while at the same time the magnetic field lines that comprise these structures are subject to Richardson-type dispersion over longer trace distances (Section~\ref{sec:disp}). In this section, we consider whether locally coherent flux ropes or dispersion of field lines governs the reconnection rate and why. In other words, does a 3D extension of 2D plasmoid mediated reconnection operate or does the LV99 theory based on field line wandering apply instead?

\subsection{Microscopic Current Layers}\label{sec:horizon:microscopic}

We prepare by recapping some principles from 2D plasmoid mediated reconnection (also see Appendix~\ref{app:pm}). Here, magnetic islands subdivide the global reconnection layer into many short reconnecting current sheets of typical half-length $l$. The microscopic current layers are on average marginally stable, such that
\begin{equation}
    S_l = \frac{V_A l}{\eta} \approx S_c,\label{eq:l}
\end{equation}
where $S_c$ is a critical Lundquist number. 
In resistive MHD, $S_c$ is usually taken to be $10^4$ and the microscopic reconnection layers reconnect at a Sweet-Parker rate
\begin{equation}
    \frac{V_{rec}}{V_A} \approx S_c^{-1/2},
\end{equation}
which determines the global reconnection rate as 0.01 \citep{2009Cassak,2009Bhattacharjee,2010HuangBhattacharjee,2010Uzdensky}. The associated typical thickness of a microscopic current sheet is 
\begin{equation}\label{eq:d}
    d = \frac{\eta}{V_A}S_c^{1/2}.
\end{equation}
If collisionless effects are important in the microscopic current sheet, then the reconnection rate is instead 0.1 \citep{2001Birn,2016ComissoBhattacharjee}.

\subsection{Alfvén Horizon}\label{sec:horizon:concept}

Reconnection in a microscopic current sheet of half-length $l$ and half-thickness $d$ has an associated time scale
\begin{equation}\label{eq:tau}
    \tau_A = \frac{l}{V_A} = \frac{d}{V_{rec}}.
\end{equation}
The first equality defines $\tau_A$ as the time in which newly reconnected plasma is expelled from the microscopic current sheet in an Alfvénic outflow jet. The second equality shows that $\tau_A$ is also the inflow time across the microscopic current layer. We will primarily view $\tau_A$ as the lifetime of a newly reconnected plasma element until it reaches the terminus of the outflow jet. The $V_A$ in Eq~(\ref{eq:tau}) is usually taken to be the Alfvén speed upstream of the reconnection layer based on the strength of the reconnecting magnetic field component.

We now introduce a causality argument that since signals travel at a finite speed, there is a horizon beyond which a plasma element cannot exchange forces during a finite lifetime. Here, we consider Lorentz forces where Alfvén waves are the force carrier. Alfvén waves are considered rather than fast magnetoacoustic waves because we are interested in sensing along the magnetic field line.

The furthest a signal can travel at the Alfvén speed during the lifetime of newly reconnected magnetic flux is $\lambda_\parallel = v_A \tau_A$, where $v_A$ denotes the average of the local Alfvén speed along the corresponding field line segment. In cases where $v_A\sim V_A$, one has $\lambda_\parallel = v_A \tau_A \sim V_A \tau_A = l$, thus the half-length of the microscopic current layer provides an upper bound on the distance out to which reconnected magnetic flux can exchange forces during its lifetime. We refer this limit on how far along the magnetic field a reconnecting plasma element can perceive its surroundings as the Alfvén horizon.

{Related causality and period matching arguments have previously been used elsewhere in plasma physics. A common application is that simulations are often designed so that plasma near the centre of the domain does not perceive the boundaries by placing the boundaries outside a causality horizon. For a recent example that focuses on Alfvén waves as the force carrier, we highlight the discussion around Eq~(46) in \citet{2019Hillier}.}

{The period matching between parallel and trnsverse dynamics that forms the second step above is also closely related to critical balance in turbulence theory \citep{1995GS} but with the modifications that the relevant perpendicular speed and length are the reconnection outflow speed $V_A$ and the jet length $l$ (which is distinguished from the microscopic current sheet thickness $d$). After submission of this paper, we became aware of independent work by \citet{2020Zhou} who used similar principles to estimate parallel scales in a different system featuring reconnection dynamics. The $R$ in their Eq~(2.4) can be interpreted as the half length of current sheet, which makes their choice of time scale identical to ours, and we remark that they also presented numerical support for these arguments. Additionally, their analysis incorporated the possibility that the reconnecting magnetic field component may be significantly weaker than the total field strength. Accounting for this increases the Alfvén horizon distance to $\lambda_\parallel = (v_A/V_A) l = (B/B_\perp) l$.}

\subsection{Coherence Dominated SGTR Simulation}

To establish the value of the Alfvén horizon in turbulent reconnection theory, we return to the SGTR simulation snapshot shown in Figure~\ref{fig:beg}. The 3D simulations by \citet{2011Daughton}, \citet{2014Daughton}, \citet{2016HB} and BRH22 all showed that locally coherent flux ropes subdivide the global current layer on horizontal planes. Inspecting Figure~\ref{fig:beg}, we focus on $z=-0.5$ and the dominant short current layer indicated by the crossed arrows in which $V_x$ changes sign. The short current layer is bookended by island-like topological features that are much thicker than the microscopic current layer (measured in the $y$ direction) and these island-like features are threaded by the green and orange flux ropes. The horizontal arrows indicate the length of the microscopic current layer on $z=-0.5$, which is consistent with Eq~(\ref{eq:l}), as we would expect for a marginally stable current sheet fragment within the turbulent reconnection layer. The $V_x$ and other quantities on $z=-0.5$ are shown in Figure~19 of BRH22.

The radius of the black circle in Figure~\ref{fig:beg} indicates the Alfvén horizon distance for the BRH22 simulation (making the $v_A \sim V_A$ simplification). The flux ropes (colored field line bundles) are subject to very little dispersion over this distance. It is therefore consistent to conclude that a reconnecting plasma element perceives the large scale magnetic field as highly coherent and hence that local coherence governs the global reconnection in this simulation.

The locally coherent structure consists of flux ropes that subdivide the global current layer into short marginally stable sections by principles familiar from 2D plasmoid mediated reconnection. The reconnection rate of these current layer fragments then sets the global reconnection rate. In resistive MHD, this rate is $S_c^{-1/2}\sim 0.01$, which agrees the values found by with \citet{2016HB} and BRH22. If collisionless effects are significant in the current sheet fragments, then the reconnection rate will be approximately $0.1$ \citep{2001Birn,2016ComissoBhattacharjee}, which agrees with the value found by \citet{2014Daughton}.

\subsection{Competition of Anisotropies}

The competition between coherence and dispersion during reconnection can be further investigated from a theoretical perspective. Combining the concept of the Alfvén horizon with Eqs~(\ref{eq:s_star}) and (\ref{eq:s_star_kpar}), a plasma element that reconnects in a microscopic current layer of half-length $l$ perceives the surrounding magnetic field as coherent on transverse scales $\delta_0$ that satisfy $s^* > \lambda_\parallel$. First considering cases where $v_A \sim V_A$,
\begin{equation}\label{eq:condition_raw}
    s^* > l 
    \quad \Leftrightarrow \quad  
    \delta_0  > (\gamma l)^{1/a}
    \quad \Leftrightarrow \quad 
    l k_{\parallel}(\delta_0) < \frac{2}{a},
\end{equation}
where $k_\parallel(\delta_0)$ signifies that $k_{||}$ for the turbulence is evaluated at the perpendicular scale matching $\delta_0$. 
 
The flux ropes that subdivide the current layer evolve over time and have a range of sizes. For 2D, a statistical theory of plasmoid sizes has been developed by \citet{2010Uzdensky}, \citet{2010Fermo,2011Fermo}, \citet{2012Loureiro} and \citet{2012HuangBhattacharjee,2013HuangBhattacharjee}. In this paper, we proceed on the simpler basis that for a flux rope to subdivide the current layer it must have grown larger than the thickness of a marginally stable microscopic reconnection layer. The most challenging coherence condition is therefore when the $\delta_0$ in inequality (\ref{eq:condition_raw}) is identified with $d$.

It follows that all flux ropes that subdivide the global current layer are effectively coherent when 
\begin{equation}\label{eq:condition_aniso}
    \frac{V_{rec}}{V_A} = \frac{d}{l} > \frac{a}{2}\left(\frac{k_\parallel}{k_\perp}\right)_{k_\perp=d^{-1}},
\end{equation}
where the subscript on the right hand side indicates that the anisotropy ratio of the turbulence is evaluated on the perpendicular scale matching the microscopic current layer thickness. We refer to inequality (\ref{eq:condition_aniso}) as the competition of anisotropies, since the magnetic field is perceived as coherent on all scales greater than or equal to the thickness of the microscopic current layer if the anisotropy of the turbulence at the perpendicular scale matching the current layer thickness exceeds the aspect ratio of the microscopic current layer.

In resistive MHD, a marginally stable Sweet-Parker reconnection layer has $V_{rec}/V_A \sim d/l \sim S_c^{-1/2} \sim 10^{-2}$. Hence, using $a=2/3$, we expect that flux-rope mediated turbulent reconnection applies when the magnetic eddies are elongated along the magnetic field direction with an anisotropy ratio $(k_\perp/k_\parallel)_{k_\perp=d^{-1}}$ greater than about 30. If a collisionless reconnection rate of 0.1 is applied instead, then turbulent reconnection will be flux-rope mediated if the anisotropy ratio exceeds approximately 3.

{Allowing for the possibility of strong guide fields modifies Eq~(\ref{eq:condition_raw}) such that $l\to(B/B_\perp)l$ and alters Eq~(\ref{eq:condition_aniso}) such that the right hand side is multiplied by $B/B_\perp$. Thus, a strong guide field increases the $(k_\perp/k_\parallel)_{k_\perp=d^{-1}}$ required for flux-rope mediated turbulent reconnection by a factor $B/B_\perp$. However, as explained in the next section, the anisotropy of the turbulence is also expected to increase with increasing guide field, such that reconnection remains flux-rope mediated.}

\subsection{{Critical Balance}}

Finally, an interesting possibility is that critically balanced turbulence generated within the global reconnection layer maintains the system in the flux-rope mediated regime. 

Suppose that the global current layer fragments into sections of half length $l$ and thickness $d\ll l$, and that the associated velocity and/or magnetic fields are responsible for forcing the turbulence. The appropriate perpendicular wavenumber would be $k_\perp \approx d^{-1}$ (since $d^{-1}\gg l^{-1}$). Meanwhile, propagation of signals along the magnetic field at the Alfvén speed would give the turbulence at this perpendicular scale an associated $k_\parallel = \lambda_\parallel^{-1} = (v_A \tau_A)^{-1}$. This proposition is supported by the recent numerical simulations by \citet{2020Zhou}. In this scenario, Eq~(\ref{eq:condition_aniso}) requires only that $a<2$, which is satisfied by most MHD turbulence scaling models including the \citet{1995GS} scaling $a=2/3$ used by LV99. 

{There is a simple physical explanation behind this conclusion. Suppose that turbulence is in critical balance, such that the parallel time scale $v_A / \lambda_\parallel$ matches the perpendicular time scale $v_\perp / \lambda_\perp$. Using causality principles from Section~\ref{sec:horizon:concept}, dynamic processes on this time scale perceive the field out to a distance $\lambda_\parallel$, i.e. a single eddy. However, as shown in Section  \ref{sec:disp:local}, the magnetic field lines remain coherent to this distance because efficient field line dispersion requires the compounding of several eddies. Thus, dynamics in a critically balanced turbulent system driven by reconnection ought to be most strongly influenced by the local coherence of the the magnetic field instead of the field line dispersion.}

\subsection{Section Summary}

This section has defined the Alfvén horizon as the maximum distance out to which plasma can exchange Lorentz forces during a time interval of interest, using Alfvén waves as the force carrier. Applied to a microscopic current layer, the maximum distance to which newly a reconnected plasma element can sense the structure of the surrounding magnetic field before the plasma element is absorbed at the terminus of the outflow is approximately equal to the half-length of the microscopic current layer (multiplied by $B/B_\perp$ if there is a strong guide field). Inspection of an SGTR simulation snapshot and theoretical analysis both concluded that magnetic flux ropes inside the turbulent reconnection layer are coherent to this distance. Thus coherence governs the global reconnection process rather than field line dispersion. 

In general, coherence dominates if the turbulence is sufficiently anisotropic at the perpendicular scale matching the microscopic current layer's thickness. Finally, we have conjectured that critically balanced turbulence generated inside the global reconnection layer may automatically satisfy this condition, since plasma percevies a distance along the magnetic field corresponding to a single eddy, and local coherence of magnetic field structure only transitions to Richardson-type dispersion once the effects of several eddies have been compounded.

\section{Discussion}\label{sec:disc}

\subsection{Central Ideas and Relation to Previous Works}

This paper has assembled a collection of principles and tools that we believe are useful for understanding magnetic reconnection with turbulence. It is motivated by recent simulations of self-generated turbulent reconnection, which exhibit features of both the LV99 theory of turbulent reconnection (large fluctuations, power law spectra, wandering of magnetic field lines and fast reconnection) and 2D plasmoid mediated reconnection (flux ropes and a reconnection rate of 0.01 for resistive MHD and 0.1 for collisionless physics). The main goal of this paper is to develop a theoretical framework for navigating the apparent duality that arises from the simulations, and hence to determine whether the global reconnection rate is governed by field line dispersion or locally coherent flux ropes.

This article and LV99 share the end goal of understanding how magnetic reconnection is affected by turbulence. Both papers also start by solving the same governing equation for the field line separation. However, the analyses diverge at an early stage for the simple reason that we included a constant of integration that was not included in  LV99. Our fuller solution for the mean square field line separation gives a theoretical basis to the observation from BRH22 that turbulent reconnection layers contain magnetic field structures that are coherent for shorter trace distances and subject to dispersion for longer trace distances. Our analytic solution also shows that the transition between these regimes occurs when the field line trace distance exceeds a few times the parallel scale of the turbulence, evaluated at the perpendicular scale matching the initial field line separation.

Another important difference between this article and LV99 is that we have addressed the conservation and conversion of magnetic helicity. When a magnetic field with non-zero magnetic helicity reconnects, some helicity must be stored inside the reconnection layer. Helicity conservation laws produce locally coherent twisted flux ropes inside the reconnection layer, as has has been observed in SGTR simulations since the pioneering particle-in-cell study by \citet{2011Daughton}. Our new work gives the flux ropes a theoretical foundation, and since local coherence and helicity laws are fundamental, it strengthens the case that the flux ropes are a robust physical feature that cannot be eliminated by increasing the Lundquist number or making the system more turbulent. The only way that we currently see to remove them while conserving magnetic helicity is to reduce the absolute value of the helicity towards zero, as in the important case where the magnetic shear angle approaches 180 degrees.

The final step was to determine whether the global reconnection process is governed by the locally coherent flux ropes or the wandering of magnetic field lines. The importance of this question can be appreciated by inspecting visualizations of the magnetic field inside the SGTR layer, such as Figure~1 of \citet{2011Daughton} or our Figure~\ref{fig:beg} (adapted from BRH22). To the best of our knowledge we are the first to formulate this question and attempt to address it using theoretical tools. Indeed, our solution for the mean square separation of field line pairs appears to be an important prerequisite to make this question approachable.

To address that final step, we transformed the question of coherence vs. dispersion into a more specific one of how a plasma element perceives the magnetic field around it, during the finite lifetime from when it reconnects in a microscopic current sheet until it is absorbed at the terminus of the microscopic outflow jet. 
Here, the central idea is that a finite time restricts interactions to a local neighborhood because signals travel at a finite speed. Considering Alfvén waves as the force carriers transmitting Lorentz forces along the magnetic field, a plasma element in a microscopic reconnection outflow jet can only exchange forces with other plasma elements inside a horizon distance approximately equal to the length of the jet (multiplied by a factor $B/B_\perp$ in the strong guide field case). Inspecting the SGTR simulation snapshot from BRH22 shown in Figure~\ref{fig:beg}, the flux ropes in the simulation are highly coherent over this distance. The Richardson-type dispersion only takes over for greater trace distances, so the plasma element cannot sense the more distant field line wandering.

Investigating theoretically, we obtained a simple and intuitive condition for coherence dominance. The transition between coherence and Richardson-type dispersion occurs at a trace distance equal to a few times the parallel scale of the turbulence, evaluated at the perpendicular scale matching the initial field line separation. Flux ropes that subdivide a global current layer must be at least as thick as the microscopic current sheets, so we adopted this thickness as the shortest perpendicular scale of interest. Then, using the Alfvén horizon, it follows that reconnecting plasma perceives the surrounding magnetic field as coherent if the parallel scale of the turbulence (at the perpendicular scale matching the thickness of the microscopic current layer) exceeds the half length of the microscopic current layer (times $B/B_\perp$ if there is a strong guide field). This can be formulated as a competition of anisotropies, where local coherence governs the reconnection if the anisotropy of the turbulence $(k_\perp/k_\parallel)_{k_\perp=d^{-1}}$ exceeds the aspect ratio of the microscopic current layer $l/d\sim S_c^{1/2}$. We find it interesting that the anisotropy of the turbulence is the important property, rather than e.g. the amplitude. Finally, we have conjectured that critically balanced turbulence generated during reconnection at high Lundquist numbers marginally maintains the flux-rope mediated regime.

\subsection{Flux-Rope Mediated Turbulent Reconnection}

The outcome is a regime of magnetic reconnection that we refer to as flux-rope mediated turbulent reconnection. Its governing prinicple is that locally coherent flux ropes subdivide the global current layer into marginally stable sections that reconnect at a fast rate that is independent of the Lundquist number. Since this principle is shared with 2D plasmoid mediated reconnection, the same reconnection rate is found in 3D and 2D. Thus, $V_{rec}/V_A \sim 0.01$ in resistive MHD \citep{2010HuangBhattacharjee,2010Uzdensky} agreeing with the SGTR simulation results of \citet{2016HB} and BRH22, and a higher rate $V_{rec}/V_A \sim 0.1$ is expected when collisionless physics is significant at small scales \citep[e.g.][]{2001Birn, 2016ComissoBhattacharjee} agreeing with the SGTR simulation results of \citet{2014Daughton}. Finally, we remark that these reconnection rates are guideline values obtained from simulations and there is further work to be done exploring the impacts of the plasma beta and other parameters.

\subsection{Differences to 2D}\label{sec:conc:not2D}

Some of the principles and properties of 3D flux-rope mediated reconnection are shared with 2D plasmoid mediated reconnection but there are also some notable differences. A glance at Figure~\ref{fig:beg} shows that 3D flux ropes are typically oblique, differing qualitatively from the parallel flux ropes that would be the equivalent of 2.5D plasmoids. Furthermore, the field line dispersion properties summarized in Figure~\ref{fig:dispersion} imply that magnetic field lines are mixed on smaller scales inside the flux ropes and current sheets. The major difference of the internal and global structure likely has major impacts on particle acceleration and trapping, as has been explored previously by \citet{2015Dahlin,2017Dahlin}. 

It has also been found that the global reconnection outflow is more diffused in 3D than in 2D, e.g. Figure~5 of \citet{2016HB}. This is consistent with our model since flux ropes are expelled from the global current layer on a time scale $L/V_A \gg \tau_A$, yielding an Alfvén horizon for the global reconnection layer of $L$. This is large enough for the field line dispersion to affect the global outflow, even though local coherence governs the reconnection rate since it dominates at the microscopic scale.

Given these differences, we believe that the 3D regime is deserving of its own name and propose it is referred to as ``flux-rope mediated turbulent reconnection''. This name is intended to convey the key similarity that reconnection is fast because the global current layer is subdivided into marginally stable fragments, while also signaling that one deals with 3D flux ropes and turbulence, which sets it apart from 2D plasmoid mediated reconnection.

\subsection{Limits of Applicability}\label{sec:disc:limits}

Non-zero helicity is required to drive the production of flux ropes with a dominant handedness inside the turbulent reconnection layer, which subdivide the global current sheet into short fragments that reconnect at a fast rate. If reconnection occurs with a 180 degree shear angle (i.e. the magnetic fields at opposite sides of the reconnection layer are anti-parallel) then the magnetic helicity is zero and the flux-rope mediated model would not apply in its current form.

Discovering what happens in the anti-parallel case requires new numerical simulations, which we are separately undertaking. The most obvious possible outcome is that field line dispersion may dominate, yielding LV99 turbulent reconnection. If so, it will be important to identify the boundary in parameter space that separates the LV99 and flux-rope mediated regimes. However, that outcome is not a given, since zero magnetic helicity is compatible with the production of flux ropes inside the reconnection layer if similar numbers are created with opposite handedness. Thus, it is not yet excluded that some form of flux-rope mediated turbulent reconnection may also occur for anti-parallel reconnection. A final complication is that the anti-parallel case permits the production of 3D magnetic null points, separatrix surfaces and separators inside the turbulent reconnection layer, so a 3D magnetic skeleton could play an important role.

The other important condition is that local coherence dominates over field line dispersion. The reconnection rates obtained by \citet{2014Daughton}, \citet{2016HB} and BRH22 are strong evidence that this condition is met in collisionless and MHD simulations in which turbulence is self-generated inside the reconnection layer. The present paper has strengthened this conclusion by showing theoretically that coherence governs the reconnection if the turbulence at small scales is sufficiently anisotropic. We remark that the condition stated in inequality (\ref{eq:condition_aniso}) is sufficient but not necessary for coherence to dominate, as the global current layer may be subdivided by flux ropes that have radii larger than the thickness of the current sheet fragments. Finally, we have conjectured that turbulent reconnection may produce a critically balanced turbulence that makes it flux-rope mediated. Recent simulations by \cite{2020Zhou} appear to support this scenario, but it is also an interesting avenue for further investigation.

Finally, if the turbulence were to have a sufficiently short parallel scale at the perpendicular scale $d=\eta S_c^{1/2}/V_A$, then the Richardson-type dispersion would presumably govern the reconnection instead. In this case, a likely outcome would be that the system behaves in accordance with the LV99 theory and resembles the driven turbulent reconnection simulations by \citet{2009Kowal} who obtained reconnection rates up to $\sim0.1$ in resistive MHD. It is interesting to ask if there is some part of parameter space in which turbulence at small scales is suficiently isotropic for this to occur. SGTR simulations produced by many groups appear to be flux-rope mediated and our analysis suggests that this is likely to be a robust result, provided the reconnecting magnetic field has non-zero magnetic helicity. Nonetheless, this transition may occur if a different mechanism forces the turbulence, e.g. if reconnection occurs in the presence of background turbulence.

\subsection{Future Questions}\label{sec:disc:open}

The flux-rope mediated turbulent reconnection model that we have presented motivates significant amounts of future work. A goal that arises immediately is to use high-resolution 3D simulations to validate key parts of the model. This is numerically challenging because the global Lundquist number $S_L$ must be much larger than $S_c$ to obtain self-generated turbulence and one must resolve the thickness of a marginally stable current sheet fragment, i.e. there is a scale separation $L/d = S_L S_c^{-1/2}$ where $L$ is the half-length of the global reconnection layer. The present work therefore motivates advancing the effective resolution of SGTR simulations as far as possible.

Next we pose the question, ``What other turbulent reconnection regimes exist and where are the boundaries in parameter space?'' Our discussion in Section~\ref{sec:disc:limits} has emphasized the likelihood that the reconnection process changes as the shear angle approaches 180 degrees, and we plan to investigate this case in the near future.

Finally, the flux-rope mediated turbulent reconnection model motivates mathematical research into identification of frayed flux ropes in 3D magnetic field data. This is much harder than the solved task of identifying plasmoids in 2D magnetic field data. Many investigators presently use current density or fluid pressure as indicators and confirm by visually inspecting field lines. However, it is desirable to develop rigorous direct methods based the magnetic field itself. Knowledge transfer from the identification of vortices with finite lifetimes in 2D fluid turbulence may be beneficial here, as may be exploring the dispersion and winding of field line pairs.

\section{Summary}\label{sec:summary}

This article has developed new tools for understanding magnetic reconnection in the presence of turbulence, emphasizing the importance of locally coherent magnetic field structures. The work provides a theoretical basis for extending certain principles established for 2D plasmoid mediated reconnection to 3D reconnection with turbulence, with locally coherent 3D flux ropes subdividing the global current layer into marginally stable fragments that reconnect at a fast rate. The key findings are as follows:
\begin{enumerate}
    \item The governing equations that \citet{1999LV} used to derive Richardson-type dispersion of magnetic field lines in turbulent plasma also imply that magnetic field structures are locally coherent. Magnetic field structures are coherent over shorter trace distances, whereas dispersion takes over for longer trace distances. The transition occurs when the trace distance exceeds a few times the parallel scale of the turbulence evaluated at the perpendicular scale matching the initial field line separation.
    \item When the reconnecting magnetic fields have non-zero magnetic helicity, twisted flux ropes with a dominant handedness form inside the turbulent reconnection layer to conserve the helicity. The most common interactions between locally coherent flux ropes are expected to be merging, which helps locally coherent flux ropes grow to a considerable size, and bouncing, which enables coexistence of oblique flux ropes on opposite sides of the reconnection layer. Unlike in 2D, the flux ropes fray for longer trace distances and magnetic field lines are mixed on smaller scales inside them. This difference is likely to have major impacts on particle acceleration and trapping, as investigated previously by \citet{2015Dahlin,2017Dahlin}.
    \item A plasma element in a microscopic current sheet has a finite lifetime that starts when it reconnects and ends when it reaches the terminus of the outflow jet. In this time interval, the plasma element can only exchange forces with other plasma elements inside a causality horizon that we have referred to as the ``Alfvén horizon''. Simulations and theoretical considerations show that the magnetic field structure is highly coherent for this trace distance, which allows subdivision of the turbulent reconnection layer by locally coherent flux ropes. This results in fast reconnection rates of 0.01 in resistive MHD and 0.1 in collisionless models.
    \item In general, coherence governs the reconnection rate if the anisotropy of the turbulence at the perpendicular scale matching the current layer thickness, $(k_\perp/k_\parallel)_{k_\perp=d^{-1}}$, exceeds the aspect ratio of the microscopic current layer (multiplied by $B/B_\perp$ for strong guide field cases). Finally, we have conjectured that turbulence generated within the global reconnection layer may automatically maintain the system in the flux-rope mediated regime, on the basis that dynamic processes in critically balanced turbulence perceive a local neighborhood corresponding to a single eddy, but local coherence only transitions to field line dispersion once the effects of several eddies have been compounded. These arguments suggest that flux-rope mediated reconnection should be widespread in solar, space and astrophysical plasmas.
\end{enumerate}

\section*{Acknowledgments}
{The authors thank the anonymous referee for constructive comments that considerably improved the paper. AJBR gratefully acknowledges funding from STFC Consolidated Grant ST/W001195/1 and valuable conversations with Gunnar Hornig, Raheem Beg, Andrew N. Wright, Andrew Hillier and Nuno F. Loureiro.} The research made use of the NRL Plasma Formulary and NASA’s ADS Bibliographic Services.

\appendix

\section{Previous Fast Reconnection Models for MHD Current Layers}\label{app}
This paper builds on two previous MHD models for fast reconnection in current layer geometries: the LV99 model of 3D turbulent reconnection \citep{1999LV}, and 2.5D plasmoid mediated reconnection \citep{2009Cassak,2009Bhattacharjee,2010HuangBhattacharjee,2010Uzdensky}. These theories are summarized here for convenience. The presentation in Section~\ref{app:LV} also demonstrates how the mathematical approach used in Section~\ref{sec:disp} can be used to solve the LV99 model instead of using scaling arguments.

\subsection{Plasmoid Mediated Reconnection}\label{app:pm}

Plasmoid mediated reconnection \citep{2009Cassak,2009Bhattacharjee,2010HuangBhattacharjee,2010Uzdensky} examines magnetic reconnection in a statistical steady state produced by plasmoid instabilities. In this 2.5D scenario, an initial reconnecting current sheet with $S_L = V_A L/\eta \gg S_c$ is tearing unstable, so magnetic islands develop in the current layer. Once the instability reaches its nonlinear stage, the current layer is fragmented into a chain of magnetic islands and current sheet segments.

Plasmoid instabilities operate recursively \citep{2001ShibataTanuma} until current sheet segments have $S_l<S_c$.  At the same time, coalescence and expulsion act to reduce the number of plasmoids in the current sheet, lengthening current sheet segments. Balancing the growth and fragmentation processes implies that a statistically stationary state contains current sheet segments with $S_l\approx S_c$.

In this 2.5D scenario, the global reconnection rate is determined by the fastest local process. In an MHD model, the current sheet segments are Sweet-Parker reconnection sites with half length $l$ satisfying $S_l\approx S_c$, hence,
\begin{equation}
    \frac{V_{rec}}{V_A} = S_c^{-1/2} \approx 0.01.
\end{equation}

The MHD model predicts that the current sheet segments have thickness $a$ such that 
\begin{equation}
    S_a = S_c^{1/2}.
\end{equation}
If $a$ is smaller than the ion skin depth $\delta_i$ or the ion Larmor radius, then collisionless physics becomes important in the microscopic reconnection layers. Simulations have shown that 
\begin{equation}
    \frac{V_{rec}}{V_A} \approx 0.1,
\end{equation}
is a robust result for a variety of collisionless models \citep{2016ComissoBhattacharjee}.

\subsection{Lazarian-Vishniac Turbulent Reconnection}\label{app:LV}

\citet{1999LV} considered the effect of a weakly stochastic magnetic field on magnetic reconnection. In their model, magnetic field lines are aligned predominantly along an $x$ coordinate parallel to the current sheet and stochastically diffused in the $y$ direction perpendicular to the current sheet. 

The key idea is that the stochastic wandering sets the effective outflow width $\Delta$. Thus, assuming conservation of mass with $\rho_{out}\approx\rho_{in}$, the normalized global reconnection rate is
\begin{equation}
    \frac{V_{rec}}{V_A} = \frac{\Delta}{L} = \frac{\left<y^2\right>^{1/2}(L)}{L},
\end{equation}
where $L$ is the half-length of the reconnection layer and $\left<y^2\right>^{1/2}(L)$ denotes the rms distance to which field lines are scattered when traced a distance $L$ in the $x$-direction. Since $\left<y^2\right>^{1/2}(L)$ is independent of the resistivity, so is the global reconnection rate.

Quantitative values for $V_{rec}/V_A$ are obtained by evaluating $\left<y^2\right>^{1/2}(L)$. Here, we present the analysis in terms of solutions to differential equations subject to initial conditions, since this method is used in the main part of the current paper, rather than the scaling analysis approach of LV99. 

LV99 considered turbulence with an injection scale $l$, and when the field line trace distance $x$ along the current sheet satisfies $x\leq l$, one can use a general scaling
\begin{equation}
    k_\parallel l = \left(\frac{v_l}{V_A}\right)^m(k_\perp l)^p,
\end{equation}
as considered in Appendix D of LV99. For Goldreich \& Sridhar scalings, $m=4/3$ and $p=2/3$. Making the identification $k_\perp\equiv\left<y^2\right>^{-1/2}$,
\begin{equation}\label{eq:LV:ODE}
    \frac{d\left<y^2\right>}{dx} = k_\parallel \left<y^2\right> = l^{p-1}\left(\frac{v_l}{V_A}\right)^m\left<y^2\right>^{1-p/2}, \quad x\leq l.
\end{equation}
This is a separable ODE with solution 
\begin{equation}\label{eq:LV:y}
    \left<y^2\right>^{1/2} = l\left(\frac{p}{2}\right)^{1/p}\left(\frac{v_l}{V_A}\right)^{m/p}\left(\frac{x}{l}\right)^{1/p}, \quad x\leq l,
\end{equation}
where we have applied an initial condition that $\left<y^2\right>^{1/2}=0$ when $x=0$.
The corresponding reconnection rate (for current layers that are shorter than the injection scale) is
\begin{equation}
    \frac{V_{rec}}{V_A} = \left(\frac{p}{2}\right)^{1/p}\left(\frac{v_l}{V_A}\right)^{m/p}\left(\frac{L}{l}\right)^{1/p-1}, \quad L \leq l.
\end{equation}

To address cases where $L>l$, we first note that the right hand side of Eq~(\ref{eq:LV:ODE}) increases with $\left<y^2\right>$ (assuming $p<2$), and Eq~(\ref{eq:LV:y}) gives $\left<y^2\right>^{1/2} = l(p/2)^{1/p}\left(v_l/V_A\right)^{m/p}$ when $x=l$. These observations yield a maximal field line diffusion rate, obtained for $x = l$, which is also applied for $x>l$. Thus, one solves
\begin{equation}\label{eq:LV:ODE_maximal}
    \frac{d\left<y^2\right>}{dx} = l\left(\frac{p}{2}\right)^{2/p-1}\left(\frac{v_l}{V_A}\right)^{2m/p}, \quad x>l,
\end{equation}
which can be integrated directly. Applying the initial condition that $\left<y^2\right>^{1/2} = l(p/2)^{1/p}\left(v_l/V_A\right)^{m/p}$ when $x=l$, the solution is
\begin{equation}\label{eq:LV:brownian_extra}
    \left<y^2\right>^{1/2} = l\left(\frac{p}{2}\right)^{1/p}\left(\frac{v_l}{V_A}\right)^{m/p}\sqrt{1+\frac{2}{p}\left(\frac{x}{l}-1\right)}, \quad x> l,
\end{equation}
which gives
\begin{equation}\label{eq:LV:brownian_extra_rate}
    \frac{V_{rec}}{V_A} = \left(\frac{p}{2}\right)^{1/p}\left(\frac{v_l}{V_A}\right)^{m/p}\left(\frac{L}{l}\right)^{-1}\sqrt{1+\frac{2}{p}\left(\frac{L}{l}-1\right)}, \quad L> l.
\end{equation}

Our solution for $x>l$ differs from the original working in LV99 because we have applied an initial condition for $x=l$ that ensures the solutions for $\left<y^2\right>^{1/2}$ are continuous and differentiable at $x=l$. However, in the limit $x \gg l$, Eq~(\ref{eq:LV:brownian_extra}) simplifies to a Brownian diffusion and Eq~(\ref{eq:LV:brownian_extra_rate}) simplifies to
\begin{equation}
    \frac{V_{rec}}{V_A} = \left(\frac{p}{2}\right)^{1/p-1/2}\left(\frac{v_l}{V_A}\right)^{m/p}\left(\frac{L}{l}\right)^{-1/2}, \quad L\gg l,
\end{equation}
which recovers the $V_{rec}/V_A \sim (v_l/V_A)^{m/p}(L/l)^{-1/2}$ scaling that LV99 applied for $L>l$.

\bibliography{TurbulentReco}{}
\bibliographystyle{aasjournal}

\end{document}